\documentclass[12pt,final]{iopart}

\usepackage{graphicx}

\begin{document}

\title[Multiple power-balanced laser beams for quantum-state manipulation experiments]{Generation of multiple power-balanced laser beams for quantum-state manipulation experiments with phase-stable double optical lattices}

\author{S. J. H. Petra, P. Sj{\"o}lund and A. Kastberg}

\address{Ume{\aa} University, Department of Physics, SE-90187 Ume{\aa}, Sweden}
\ead{anders.kastberg@physics.umu.se}
\begin{abstract}
We present a method to obtain power-balanced laser beams for doing quantum-state manipulation experiments with phase-stable double optical lattices. Double optical lattices are constructed using four pairs of overlapped laser beams with different frequencies. Our optical scheme provides a phase stability between the optical lattices of 5~mrad/s, and laser beams with a very clean polarisation state resulting in a power imbalance in the individual laser beams of less than 1\%.
\end{abstract}

\vspace{2pc}
\maketitle

\section{Introduction and background}
Optical lattices are periodic arrays of shallow microscopic traps in which ultra-cold atoms can be trapped \cite{Jessen:1996,Grynberg:2001}. The trapping potential is a consequence of a second-order interaction between the atomic dipole moment and an applied light field \cite{Cohen-Tannoudji:1992}: the associated energy, which is proportional to the irradiance, is generally referred to as the light shift. The periodicity emerges when the total light field is built up by two or more interfering laser beams.

Optical lattices are widely used as a tool for a range of different scientific studies. In particular, they have lately been extensively utilised in experiments involving Bose-Einstein Condensates \cite{Bloch:2005} and quantum-state manipulation \cite{Monroe:2002}. In the latter context, cold atoms in optical lattices have been suggested as a promising candidate for a platform for quantum computing.

The attractiveness of optical lattices in attempts to realise rudimentary quantum gates, and in subsequent more ambitious steps towards a quantum computer, comes mainly from the very efficient isolation from ambient effects. Early suggestions for quantum information processing with optical lattices \cite{Brennen:1999,Jaksch:1999,Sorensen:1999} do, however, impose strict, and sometimes conflicting technical demands on the optical lattice.

In this paper, we present a solution for obtaining power balanced laser beams to create phase-stable double optical lattices, which is essential in order to realise a quantum gate and to enable other classes of highly controlled quantum-state manipulation in dissipative optical lattices. We have previously presented a way to realise a double set of independent 3D optical lattices that are phase stable \cite{Ellmann:2003_PRL}. Many of the details behind the phase control and the acquired stability can be found in \cite{Ellmann:2003_EPJD}.

\subsection{Power balance}
When generating a 3D optical lattice with four (minimum) or more laser beams, it can be paramount to have equal power in all beams, especially when dealing with dissipative, near-resonant optical lattices. One example of this is when the depth and the oscillation frequency of the potential wells have to be precisely known (and reproducible) for controlled preparation of q-bits or for general investigations of the properties of optical lattices. Another example is when using the optical lattices in order to controllably induce small directed drifts in the atomic population \cite{Sjolund:2005}. In the latter case, a spurious drift due to radiation pressure caused by a power imbalance is likely to wash out the desired effect. Equal power in all beams can therefore be of more importance than overall power stability in time, since small fluctuations in the total power do not affect the relative power in the individual beams.

As long as the optical lattice is of a standard type, involving only one wavelength, obtaining equal power in all beams is trivial e.g.\ by adjusting the polarisation of each individual beam before passing through a polarising beam-splitter cube, the transmitted power can be equalised for all beams. However, in many suggestions for how to realise quantum computing with optical lattices, more than one wavelength is needed. In our set-up for double optical lattices, we use two laser colours, albeit close in wavelength, and in order to keep the spatial phases of the optical lattices stable, we have to use an elaborate scheme where light from two different lasers are overlapped with crossed polarisation \cite{Ellmann:2003_EPJD}. Eventually, the double optical lattice is formed by eight laser beams arranged in four pairs. Each pair contains two laser fields of different wavelengths but with an identical spatial mode. In this case, it is not possible to control the power of all beams individually as is done with a single wavelength, since the beams have orthogonal polarisations and are physically overlapped. Adjusting the polarisation of the beams to control the power for one frequency will therefore also change the power in the other frequency, which means that power balance for both frequencies cannot be controlled independently.

In a recent experiment, where a 3D controllable Brownian motor was demonstrated \cite{Sjolund:2005}, the rudimentary power balance in the first experiments with double optical lattices was not enough. Therefore, an optical system for controlling the power, without compromising the phase control had to be devised.

\section{Methodology}

\subsection{Double optical lattices}
A double optical lattice is two overlapped optical lattices that have identical topography, but that can be controlled individually in terms of well depths and spatial phases. The inherent difficulty is that of those two prerequisites, the former requires that the two lattices are built up by the same wavelength, whereas the latter requires individual beams of different wavelengths in order to avoid cross talk. For our double optical lattice we use two laser frequencies that are near-resonant with the D2 line of the cesium atom, which has two hyperfine ground states, separated by 9.2~GHz (c.f.\ \cite{Ellmann:2003_PRL,Ellmann:2003_EPJD}). These resonances are much narrower than the difference in the laser frequencies, which allows for independent addressing of the two lattices. At the same time, the wavelengths are close enough, and the sample volume is small enough (1--2 mm in diameter), to ensure that the periodicity is constant for practical purposes (it takes 3--4 cm for the lattices to phase out).

\subsubsection{Phase-stability and phase-control of the double optical lattice}
Since optical lattices are created in interference patterns, the resulting spatial phase of an optical lattice will depend heavily on the optical phases of all laser beams involved. When using two optical lattices built up by different wavelengths, it is crucial that this is under control. 

We choose a configuration with only four beams per 3-dimensional lattice (one beam more than the dimensionality \cite{Grynberg:1993}). This ensures time-constant topography in each individual lattice. A phase shift in any one of the four laser beams will only result in a global translation of the lattice, and that on a time scale which is typically irrelevant for the atomic dynamics in the lattice. Typical single-well oscillation frequencies in our optical lattice are of the order of hundreds of kilohertz, which is much larger than any mechanical or acoustic noise. This means that even though we generate the two colours with independent lasers, these lasers do not have to be phase locked to each other. All phase fluctuations, including those in the lasers, that happen before the laser light is split up into several branches, will appear identically in all these branches. Thus, the spatial phase of the interferogram will remain constant regardless of such fluctuations.

\begin{figure}[ht]
\begin{center}
\includegraphics[width=\textwidth,keepaspectratio]{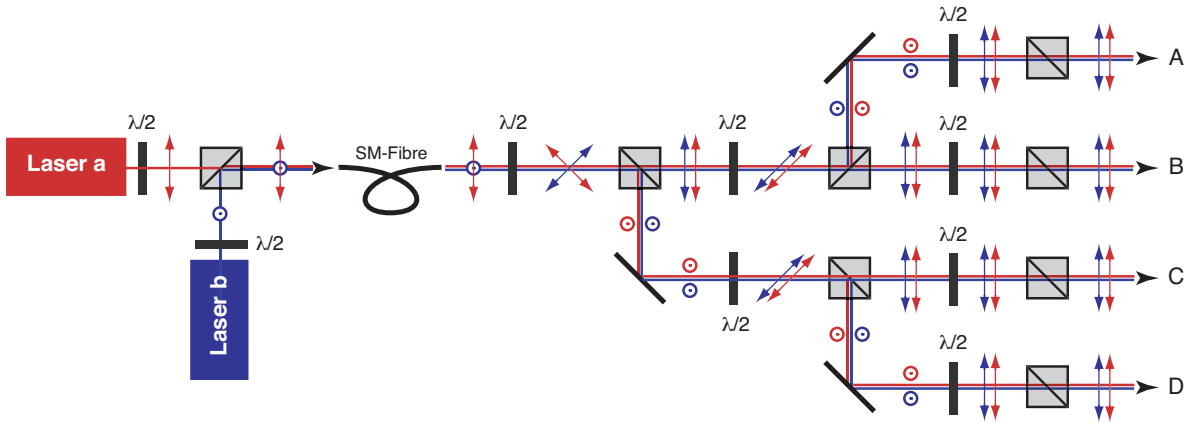}
\caption{Optical configuration for ensuring phase stability for a double optical lattice. Beams (red and blue) from two lasers, with different frequencies, and with crossed polarisations are overlapped with a polarising beam-splitter cube. They are then injected, together, into a single-mode optical fibre. They exit the fibre, still with approximately crossed polarisation. The polarisations are rotated before the pair is split up by a polarising beam-splitter. The emerging two beam pairs now each consist of two frequency components, but now with parallel polarisations within each pair. These pairs are then split up again in order to provide the needed four beams pairs, where the two frequency components in each respective branch are in the same spatial mode.
\label{fig1}}
\end{center}
\end{figure}

Phase fluctuations in an individual beam will however lead to a global translation of the lattice. Since we operate two lattices, built up by individual laser fields, the control of the relative phase between the lattices will deteriorate if this is not dealt with. We solve this by combining the two laser frequencies, with crossed polarisation, before they are split up into four branches (see figure~\ref{fig1}). To ensure that the spatial modes actually become identical, we inject this pair of beams into a single-mode optical fibre. After the output of the fibre, the crossed polarisation of the beam pair is rotated such that it becomes 45$^\circ$ with respect to the plane of incidence of a polarising beam-splitter cube. Thus, two beam pairs are obtained, where each contains two laser frequencies, but now with the same polarisation. These are then split up yet again to provide four branches of beam pairs. To ensure equal and clean polarisation in all branches, each beam pair finally passes a half-wave plate and a polarising beam-splitter cube. A phase fluctuation in an optical component, caused for example by mechanical noise, will now occur in an identical way for both optical lattices. Thus, insignificantly slow global translations still happen, but these are now synchronised between the lattices. The relative phase of the lattices remains constant, as long as the mechanical fluctuation is less than several centimetres.

The relative phase can still be manipulated. Each optical arm contains a delay-line (not indicated in figure~\ref{fig1}), consisting of two retro-reflecting prisms \cite{Ellmann:2003_EPJD}. By adjusting the distance between the two prisms, the relative phase of the optical paths can be set for each individual beam. Since the wavelengths are so close, this distance has to be changed in the order of a centimetre in order to impose a significant phase variation. The relative phase can also be manipulated fast by using electro-optical modulators.

\subsection{Polarisation control and power balance}
Despite the care taken to have clean polarisations, it is impossible to obtain good power balance in the four beams pairs with the set-up shown in figure~\ref{fig1}. First, the polarisation state of the two beams coming out of the optical fibre is not stable, and shows some ellipticity. This causes a power imbalance of more than 30\% in the four branches. Second, the polarising beam-splitter cubes we use reflect up to 5\% of the p-wave, which ought to be transmitted, while less than 0.5\% of the s-wave, which should be reflected, is transmitted. These defects introduce an asymmetry in the system that cannot be compensated for when simultaneous power balance in the four beams for both lasers is required. Compensating for the unequal power ratios in the four beams for one laser, by adjusting the half-wave plates, leads to a larger power imbalance for the other laser.

\begin{figure}[ht]
\begin{center}
\includegraphics[width=0.7\textwidth,keepaspectratio]{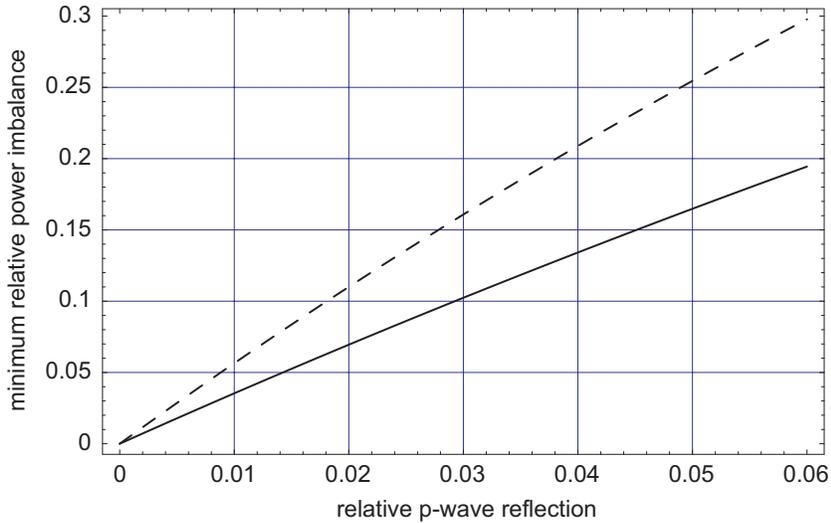}
\caption{Calculated minimum relative power imbalance in the four beams as a function of the polarising beam-splitter cube defect (the amount of light in the p-wave that is reflected) for laser A (solid line) and laser B (dashed line).
\label{fig2}}
\end{center}
\end{figure}

To quantify the power imbalance caused by the imperfect polarising beam-splitter cubes, we calculated the power in the four beams for both lasers using Jones matrix formalism \cite{Hecht:2002}. Figure~\ref{fig2} shows the calculated relative power imbalance in the four beams for both lasers as a function of the defect of the polarising beam-splitter cubes, i.e., the relative amount of light of the p-wave that is reflected rather than transmitted. The fact that the beams have a different dependency on the polarising beam-splitter cube defect makes it impossible to balance the beams for both lasers.

\begin{figure}[ht]
\begin{center}
\includegraphics[width=\textwidth,keepaspectratio]{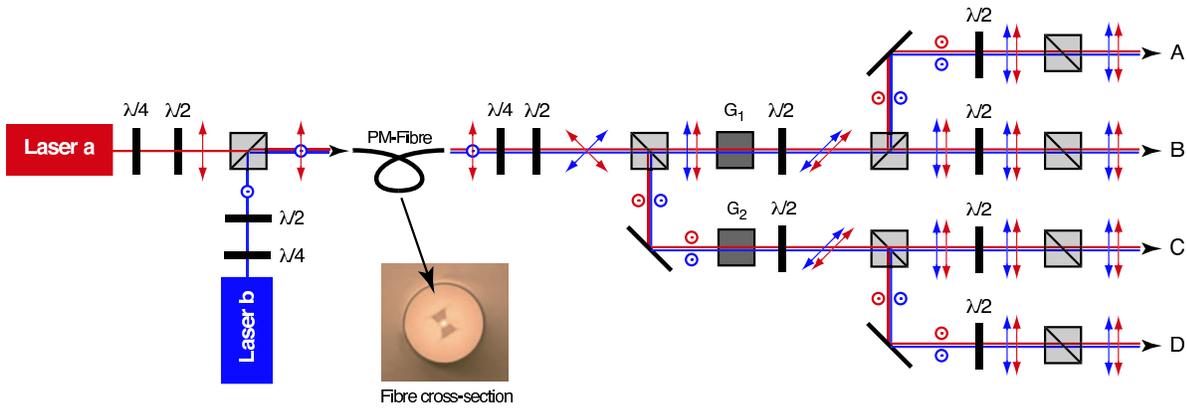}
\caption{Improved optical configuration, necessary for having power balanced beams. In order not to compromise the phase-stability, the main features of the original set-up (as depicted in figure~\ref{fig1}) must be preserved. In the improved set-up, the optical fibre has been replaces by a bow-tie stressed polarisation maintaining fibre, and great care has been taken to clean up polarisations, for example by introducing Glan-Thompson polarisers (G$_1$ and G$_2$). The photo of the fibre cross-section is courtesy of Fibercore Ltd. \label{fig3}}
\end{center}
\end{figure}

The key to obtain the necessary power balance is polarisation control. In figure~\ref{fig3}, we show how the optical set-up has been upgraded. First, when the beams are combined, it is crucial to ensure normal incidence to the first beam-splitter cube, and that ellipticities in the polarisations are eliminated. The combined beam pair is then injected to a 10~m long bow-tie stressed polarisation maintaining optical fibre \cite{Noda:1986}. This fibre (manufactured by Fibercore Ltd.) is specifically designed to maintain two different, orthogonal polarisations. The mixing of polarisation in the fibre is specified to be maximum one part in 10$^5$ per meter. Also, the output polarisation state is determined by the orientation of the fibre in the mount, and is therefore stable in time. After the output from the fibre, ellipticity is again eliminated and the polarisations of the beam pair is rotated with the respect to the subsequent polarising beam-splitter cube just as before. In each output arm of this beam-splitter cube, where the polarisations should now be perfectly linear, the polarisation state is further cleaned by installing Glan-Thompson polarisers, with a specified maximum polarisation impurity of one part in 10$^5$. For all polarising beam-splitter cubes, also after the fibre, it is important that the optical incidence is normal.

\begin{figure}[ht]
\begin{center}
\includegraphics[width=0.7\textwidth,keepaspectratio]{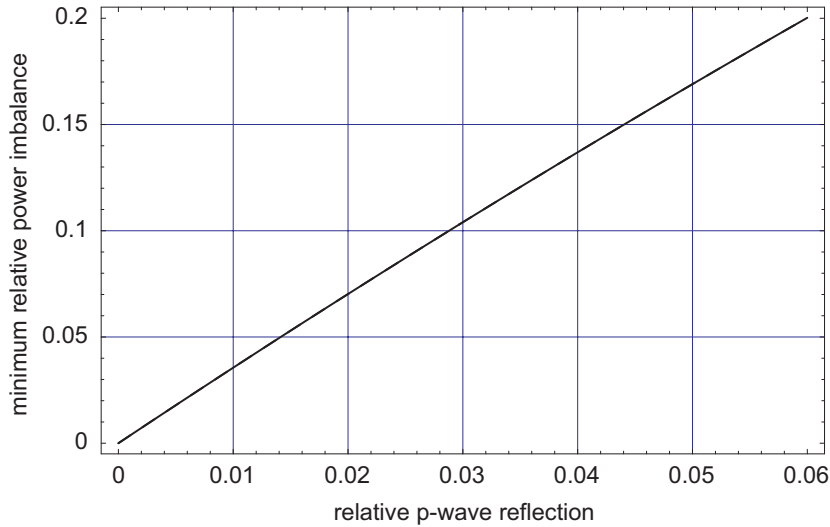}
\caption{Calculated minimum relative power imbalance in the four beams as a function of the polarising beam-splitter cube defect, for both lasers A and B, when Glan-Thompson polarisers are used to clean up the polarisation.
\label{fig4}}
\end{center}
\end{figure}

The Glan-Thompson polarisers ensure that in each of the two arms, both laser beams have exactly the same polarisation before being split-up further. With this well-defined polarisation state, it is possible to obtain power balance between all of the four branches. The effect of the Glan-Thompson polarisers on the power imbalance is depicted in figure~\ref{fig4}. In this figure, the calculated power imbalance of both lasers show the same dependency on the polarising beam-splitter cube defect; the curves for both lasers completely overlap. The power in branches A and B will be slightly less than the power in branches C and D (see figure~\ref{fig3}), due to the defect of the first beam-splitter cube after the optical fibre. However, this can now easily be compensated for by adjusting the half-wave plates in arms C and D.

\begin{figure}[ht]
\begin{center}
\includegraphics[width=0.7\textwidth,keepaspectratio]{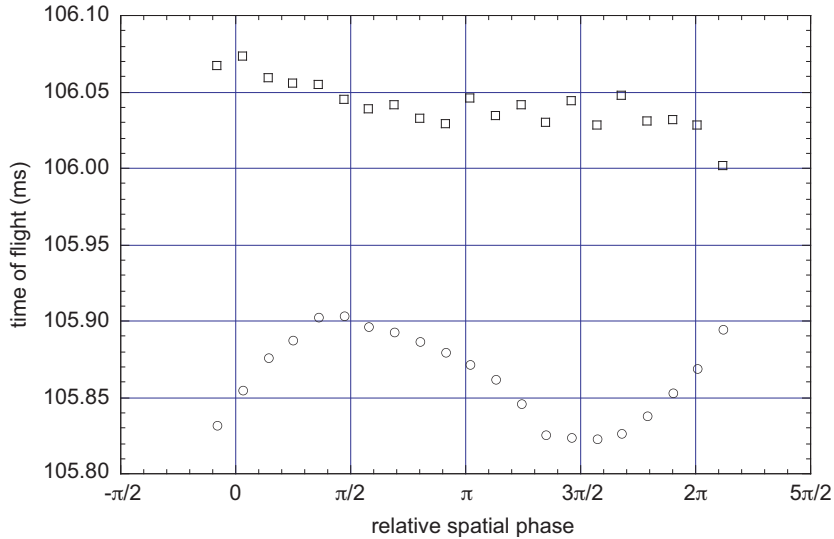}
\caption{Time-of-flight measurements of a cloud of laser-cooled atoms when released from the double optical lattice as a function of the relative spatial phase of the two lattice potentials. The experiment is performed with the old set-up (squares), with a power-imbalanced laser beams, and the modified set-up (circles), where all the beams have equal power.
\label{fig5}}
\end{center}
\end{figure}

To demonstrate the importance of having power-balanced light beams, we have performed experiments on Brownian motors, in which we measure an induced drift velocity of a cloud of cold atoms in the double optical lattice \cite{Sjolund:2005}. It can be shown that the atoms will experience an induced drift in a direction that depends on the relative spatial phase of the two optical lattices \cite{Sanchez-Palencia:2004}. An induced drift in the vertical direction will affect the time of flight of the atoms arriving at a resonant probe beam after being released from their trapping potential. Details of the experimental set-up, including the alignment of the double optical lattice, can be found in \cite{Ellmann:2003_EPJD}. The results of the experiments are shown in figure~\ref{fig5}, where the arrival time of the atoms at a probe beam is plotted as a function of the relative spatial phase of the two lattices. The experiment is done using both the old set-up (squares) and the modified set-up (circles). The data show a strong dependency of the relative phase in the upgraded set-up (i.e.\ with power balanced beams), in agreement with a theoretical model \cite{Sanchez-Palencia:2004}. The phase dependency on the arrival time cannot be seen in the old set-up. Instead, the atoms are always pushed upwards due to the radiation pressure of the power-imbalanced laser beams, resulting in a longer time of flight. From the figure it becomes clear that power-balanced laser beams are a necessity for high precision experiments, as the Brownian motor.

\section{Results and conclusions}
With the current set-up, we can maintain a phase-stability of 5~mrad/s. This stability is determined by thermal fluctuations and mechanical vibrations that cause path-length fluctuations in the individual optical paths of the laser beams. We can set the power imbalance to be maximum 1\% for both optical lattices simultaneously. The balancing is limited by the accuracy of the power meters that are used to measure the power of the beams in each branch.
 
The improvements to the double optical lattice experiments were crucial for our recent experiments on Brownian motors \cite{Sjolund:2005} and will also be in future work on quantum gate and quantum state manipulation in general. In \cite{Sjolund:2005} we can detect induced atomic drift velocities as small as 0.1~mm/s. The spurious drift caused by imbalanced power in the beams is still one of the limiting factors for this detection.

The method presented can potentially be useful also to general interferometric applications where multiple frequencies are applied with a high level of control.

\section*{Acknowledgements}
This work has been supported by Knut och Alice Wallenbergs stiftelse, Carl Tryggers stiftelse, Kempestiftelserna, Magnus Bergwalls stiftelse and the Swedish Research Council.

\section*{References}


\end{document}